\documentclass[preprint2]{emulateapj}

\slugcomment{version 2009 Jan 14: fm}

\shorttitle{Extended X-ray emission in 3C 305}
\shortauthors{Massaro F., Chiaberge M., Grandi P., et al.}

\newcommand{\hst}{{\it HST}}

\newcommand{\xmm}{{\it XMM-Newton}}
\newcommand{\chn}{{\it Chandra}}

\begin{document}


\title{Extended X-ray emission in radio galaxies: \\ the peculiar case of 3C 305}


\author{F.~Massaro\altaffilmark{1}, 
M.~Chiaberge\altaffilmark{2,4}, 
P.~Grandi\altaffilmark{3}, G.~Giovannini\altaffilmark{4.9}, C.~P.~O'Dea\altaffilmark{5}, 
F. D. Macchetto\altaffilmark{2}, S.~A.~Baum\altaffilmark{6}, R.~Gilli\altaffilmark{5}, A. Capetti\altaffilmark{8},
A.~Bonafede\altaffilmark{4.9}, E. Liuzzo\altaffilmark{4.9}}
\altaffiltext{1}{Smithsonian Astrophysical Observatory, 60 Garden Street, Cambridge, MA 02138}
\altaffiltext{2}{Space Telescope Science Institute, 3700 San Martine Drive, Baltimore, MD 21218}
\altaffiltext{3}{INAF-IASF - Istituto di Astrofisica Spaziale e fisica cosmica di Bologna, Via P. Gobetti 101, 40129, Bologna, Italy}
\altaffiltext{4}{INAF - Istituto di Radioastronomia di Bologna, via Gobetti 101 40129 Bologna, Italy}
\altaffiltext{5}{Dept of Physics, Rochester Institute of Technology, Carlson Center for Imaging Science 76-3144, 84 Lomb Memorial Dr., Rochester, NY 14623}
\altaffiltext{6}{Carlson Center for Imaging Science 76-3144, 84 Lomb Memorial Dr., Rochester, NY 14623}
\altaffiltext{7}{INAF - Osservatorio Astronomico di Bologna, Via Ranzani 1, 40127, Bologna, Italy}
\altaffiltext{8}{INAF - Osservatorio Astronomico di Torino, Strada Osservatorio 20, I-10025 Pino Torinese, Italy}
\altaffiltext{9}{Dipartimento di Astronomia, Universita' di Bologna, via Ranzani 1, 40127 Bologna, Italy}
\begin{abstract} 
Extended X-ray structures are common in Active Galactic Nuclei (AGNs).
Here we present the first case of a Compact Steep Spectrum (CSS) radio galaxy, 3C 305, in which the X-ray radiation appears to be 
associated with the optical emission line region, dominated by the [O III]5007. On the basis of a morphological study,
performed using the comparison between the X-rays, the optical and the radio band, 
we argue that the high energy emission has a thermal nature and it is not directly linked to the radio jet and hotspots of this source. 
Finally, we discuss the origin of the extended X-ray structure connected with the optical emission line region following two different interpretations:
as due to the interaction between matter outflows and shock-heated environment gas, or as due to gas photoionized by nuclear emission.
\end{abstract}

\section{Introduction}
The origin of the extended X-ray emission from Compact Steep Spectrum (CSS) radio galaxies is still unclear. 
It may be expected as a result of the radio source evolution and its expansion in the surrounding interstellar and intergalactic medium (ISM/IGM) (Heinz et al. 1998). 
It can be generally described as due to 1) a relic of the past activity, 2) a confining medium, 3) signatures of interactions
between the jet and the ISM or if present, 4) to an X-ray cluster (Siemiginowska et al. 2008).
The informations in the X-ray band are limitated by the number of available observations.

On the other hand, several analyses have been performed on the optical emission line region in CSS sources (e.g Labiano et al. 2005, Privon et al. 2008)
and there is still a considerable debate on its origin, despite the larger amount of data available in the optical band than in the X-rays. 
Narrow band HST imaging in [O II] and [O III] emission lines has shown that in these sources the optical line emission is co-spatial and strongly aligned with the 
radio source at all redshifts (e.g. De Vries et al. 1998, Axon et al. 2000, Privon et al. 2008).
This radiation can be interpreted as due either to the photoionization from the nucleus or from the jet-environment interactions, or a combination of the two. 
In the case of the jet-environment interaction scenario the origin could be
in the form of collisional ionization or a `autoionizing' shock, as suggested by Dopita \& Sutherland (1995, 1996).

The connection between the extended soft X-ray structure and the optical emission line region has been recently investigated in the case of radio-quiet AGNs.
A detailed X-ray spectral analyses of the three brightest Seyfert galaxies, NGC 1068 (Kinkhabwala et al. 2002, Brinkman et al. 2002, Ogle et al. 2003)
Circinus (Sambruna et al. 2001, Massaro et al. 2006) and Mrk 3 (Sako et al. 2000, Bianchi et al. 2005, Pounds \& Page 2005) have shown that their soft X-ray
spectrum look to be  a blend of emission lines from He- and H-like light metals and L transitions of Fe. 
These emission lines are produced by photoionized gas rather than hot gas in collisional equilibrium (Bianchi et al. 2006).
\chn~observations have shown that the soft X-ray emission of these three sources is extended (Sambruna et al. 2001, Sako et al. 2000).

Combining the high spatial resolution capabilities of \chn~ with those of Hubble Space Telescope (\hst),
Young et al. (2001) compared the extended optical emission with that in the X-ray band in the Seyfert galaxy NGC 1068.
They found that the size and the morphology of the soft X-ray region is associated with the Narrow Line Region (NLR), 
mapped by the [O III]5007 emission line, interpreting this emission as due to photoionization and fluorescence of 
gas by radiation from the Seyfert nucleus located several kpc from it.

Bianchi et al. (2006) performed a deeper investigation on a sample of 8 Seyfert galaxies observed by \chn, \xmm~and \hst~selected 
on the basis of an extended [O III]5007 emitting region.
They unambiguously showed that there is a spatial correlation between the soft X-ray emission and their [O III]5007 emitting region.
The \xmm~RGS spectral analysis and the comparison with numerical simulations also suggested that
the soft X-ray component is mainly produced by photoionized gas, and described in terms of a blending of emission lines.
They also showed that the X-ray and the [O III]$\lambda$5007 emitting gas had the same ionization source.
The role of radio jets in this scenario is still under debate and they seem to be related to the NLR/soft X-ray emission.

The hot ~0.5-1 keV normal ISM of the host elliptical is also detected in radio galaxies (O'Dea et al. 2006, Evans et al. 2005, Worrall et al. 2003).
In these sources, the association between the optical emission line region and the X-ray diffuse emission has been recently found only in few cases (e.g. 3C 321, Evans et al. 2008, 3C 33, Kraft et al. 200), where the X-ray emission has been interpreted due to gas collisional ionized. 
However, there are also few examples of radio galaxies or quasars in which the soft X-ray spectrum has been described in terms of emission lines due to photoionization, 
similar to the Seyferts, as 3C 234 (Piconcelli et al. 2008).

The radio galaxy 3C 305, observed as part of the \chn~snapshot survey 
of 3C  sources at redshift lower than 0.3, 
(see Massaro et al. 2008a and Macchetto et al. 2008, for preliminary results) has revealed a very 
intriguing morphology, characterized by an extended soft X-ray emission (0.2--2.0 keV), not 
coincident with its radio structure.

3C 305 is relatively low radio power radio galaxy located at redshift $z$ $\sim$ 0.0416, 
It has been classified as a peculiar FR I on the basis of the radio power and morphology (Heckman et al. 1982 ) and it 
presents a prominent extended emission line region of roughly the same dimension as the radio structure.

The first multiwavelength investigation of 3C 305 was carried out by Heckman et al. (1982), who presented radio maps, and optical 
images and spectroscopy of the central region. They found that the emission line region is approximately dumbbell-shaped and oriented NE-SW.
The radio emission follows this pattern, with the exception of a plume extending south ward of the NE lobe and a lack of coincidence between the radio and 
the optical structure in the lobe. It is also highly depolarized where the emission lines are brighter.

Jackson et al. (1995) presented the first \hst~ image of this radio galaxy. 
They found that the major concentration of the line emission lies beyond the end of the radio jet, 
suggesting that it can be excited by a shock propagating outwards from an 
interaction between the radio jet itself and the external medium. 
This extended optical emission line region is also associated with an outflow of neutral hydrogen (Morganti et al. 2005), 
and the [O III]5007 emission line region is aligned with the jets and hotspots (Privon et al. 2008).

A study of the neutral hydrogen absorption (H I) has been conducted on the basis of MERLIN observations by Jackson et al. (2003), in comparison with \hst~ optical and infrared images.
They argue that profiles of the [Fe II] line at 1.644 $\mu m$ and [O III]5007 are different: in the northern region, 
where the jet terminates, there is the peak of the [Fe II] line while the [O III]5007 is more evenly distributed throughout  the emission line region. 
The interpretation suggested by Jacskon et al. (2003) on the basis of the [Fe II] analysis was in favor of an emitting gas collisionally ionized by shocks around an interaction region.

In this letter we present the comparison between the radio, optical and X-ray images of 3C 305 to investigate the morphology and the nature 
of the emitting gas. For the first time in a CSS source we found that the soft X-ray emission is associated with the optical emission line region, 
dominated by the [O III], arguing that its nature is thermal.

We assumed a flat cosmology with $H_0=72$ km/(s Mpc), $\Omega_{M}=0.27$ and $\Omega_{\Lambda}=0.73$ (Spergel et al. 2007).
It means that for 3C 305, 1'' corresponds to 0.811 kpc, and its luminosity distance is 181.6 Mpc.

\section{The peculiar radio morphology}
We investigated the radio structure of 3C 305, on the basis of archival VLA and Merlin data, to derive a better classification of the source and to
compare the emission of its radio jet and hotsposts with that in the X-rays. 
 
The radio emission of 3C 305 at 1.7 GHz \footnote{Radio map available on the website: http://www.jb.man.ac.uk/, "An Atlas of DRAGNs"}, is peculiar (Heckman et al. 1982, Jackson et al. 2003). 
It is characterized by an extremely faint radio core with two-sided
jets, which terminates in a high brightness region (similar to but fainter
to hotspots which characterize powerful radio sources), from where the radio emission
extends with a plumed double structure with a very symmetric H shape.
The NE lobe is more extended to the SE with a low brightness region 
in between the core and the bright spot. The SW lobe is more symmetric with 
the bright spot located at the center of the low brightness region.

We reduced VLA archival data at 8.4 GHz with the AIPS standard reduction procedure 
(Amplitude calibrator: 1331+305, phase calibrator: 1637+574).
The final image is in good agreement with the 1.7 GHz radio map.
To derive a spectral index distribution we obtained images at 1.7 GHz and 
8.4 GHz with the same cellsize and resolution, matching the uv-coverage as
much as possible. The jets, lobes,
and also the two bright spots have a moderately steep spectrum 
($\alpha^{8.4}_{1.7} \sim$ 0.8) with no clear trends.
The only steeper regions are in the external part of the NE lobe and unexpected in the inner part of the SW lobe.

This source is small in size: only $\sim$ 3.5 kpc is the distance between
the two bright spots. The total radio spectrum is straight and steep (we found: $\alpha^{86}_{14900}$ = 0.9 from the archive data
\footnote{We assume flux density S($\nu$) $\propto$ $\nu^{-\alpha}$}) suggesting that 3C 305 is an aged confined source.
Its steep radio spectrum together with the small size extension is in agreement with the identification of 3C 305 as a CSS source.

The core is very faint in the radio band ($<$ 2 mJy at 1.7 GHz),
in agreement with Jackson et al. (2003). 
From this result and the jet to counter-jet brightness ratio we
estimated that this source is oriented near to the plane of the sky
($\sim$ 65-70 degree) and that the jets are highly relativistic in the inner
region ($\sim$ 0.9c) and slow down reaching a velocity 
$\sim$ 0.5c at $\sim$ 2" ($\sim$ 1.5 kpc) from
the core. This picture is confirmed by the marginal core detection in
VLBI data (Giovannini et al. 2005).

\begin{figure}[!htp]
\includegraphics[height=5.5cm,width=7cm,angle=0]{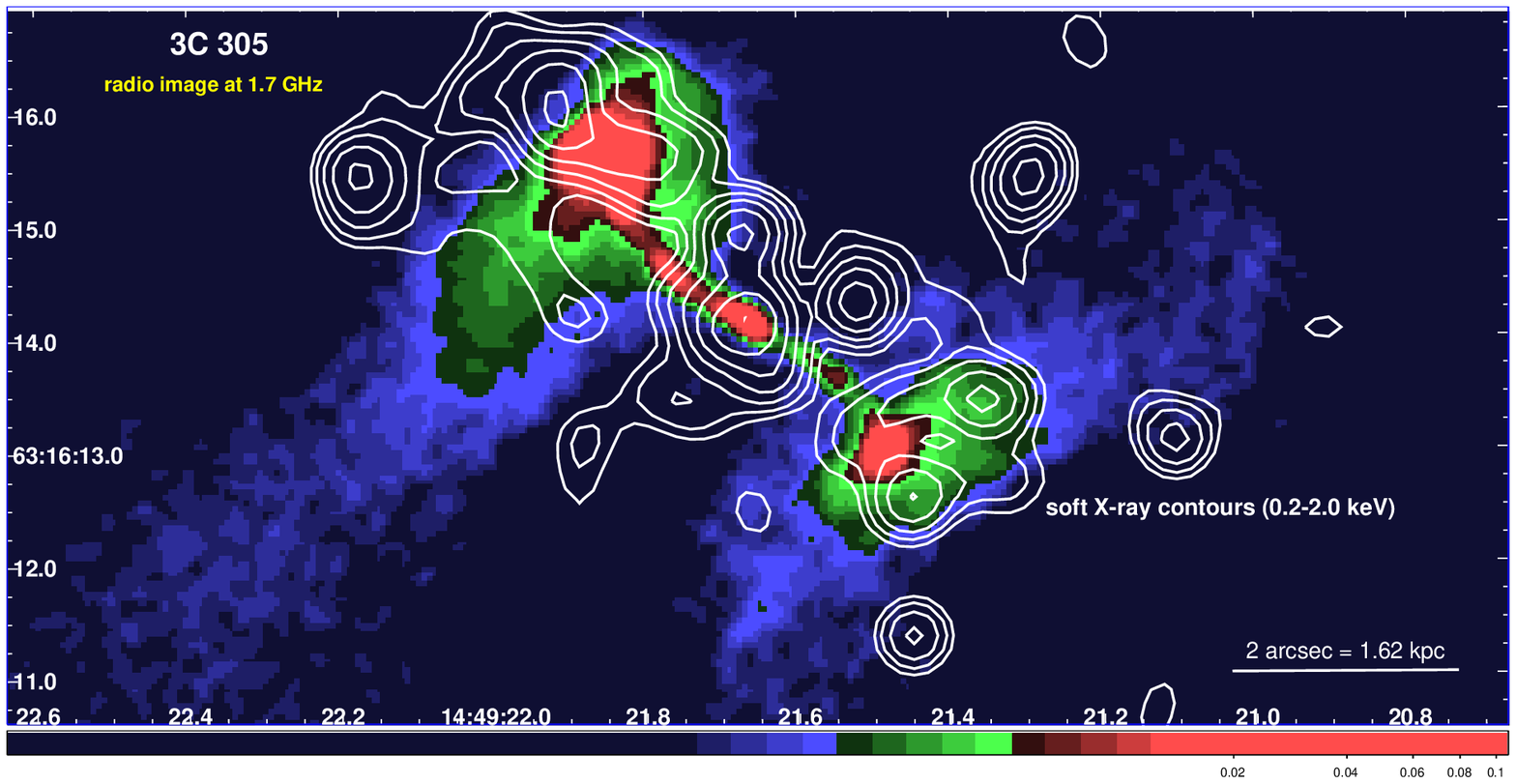}
\includegraphics[height=5.5cm,width=7cm,angle=0]{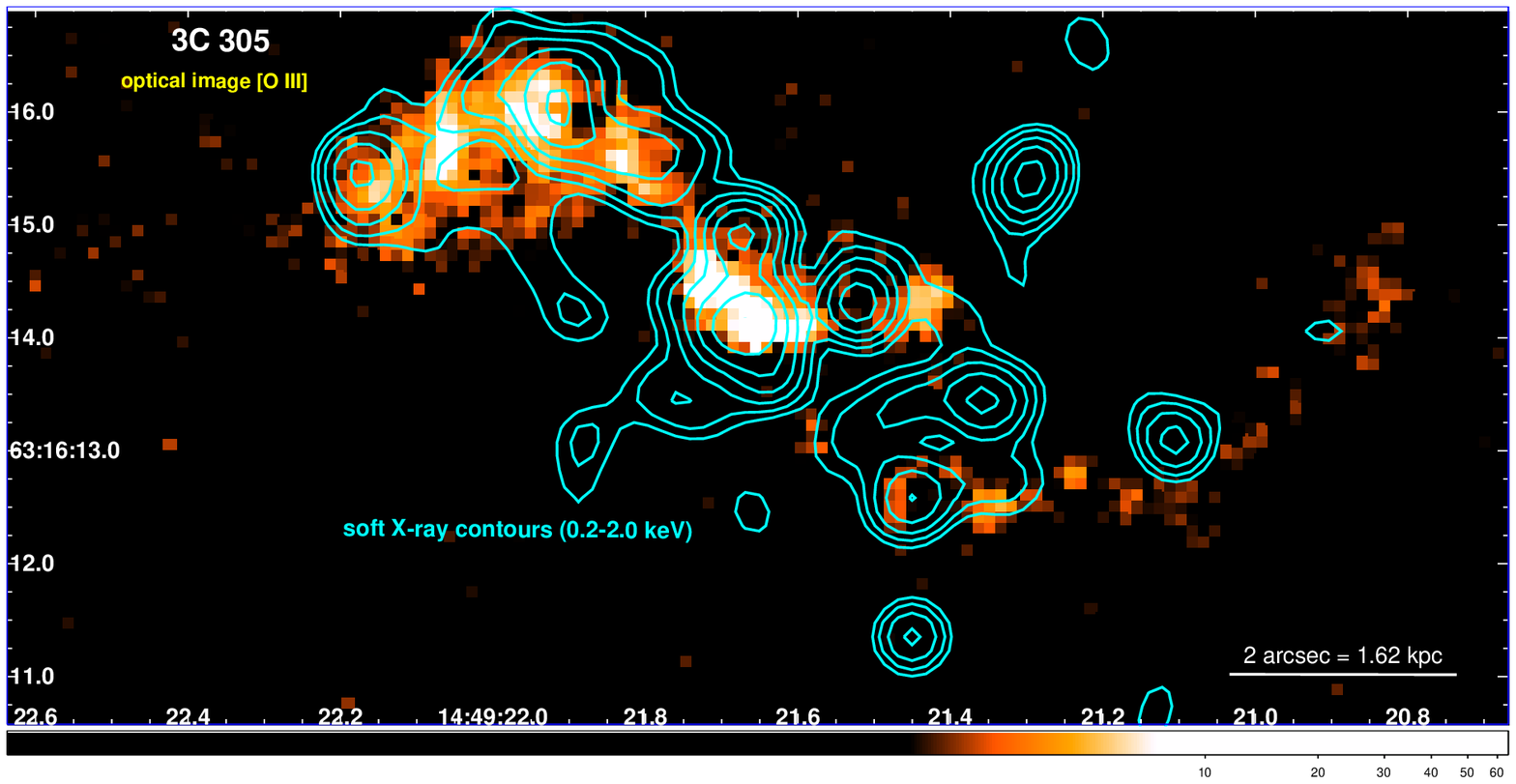}
\caption{{\it Left panel}: The radio image at 1.7 GHz obtained combining MERLIN and VLA data retrived from the website: http://www.jb.man.ac.uk/, "An Atlas of DRAGNs".
with soft X-ray contours overlaid (0.2--2.0 keV, smoothed 0.5''). There is a clear evidence that the extended X-ray emission lies beyond the radio structure. {\it Right panel}: HST [O III]5007 image, (continuum subtracted image to exclude the stellar emission, see Privon et al. 2008) 
with X-ray contours overlaid (0.2--7 keV energy range, smoothed 0.5'') (cyan). 
Here a clear evidence of the X-ray emission associated with the emission line region is shown. }
\label{fig1}
\end{figure}

\section{The extended X-ray emission}
3C 305 has been observed by Chandra (Obs ID 9330) on April 7, 2008, with the ACIS-S camera, operating in VERY FAINT mode, with an exposure of about 8 ksecs. 
The data reduction has been performed following the standard
procedures described in the Chandra Interactive Analysis of
Observations (CIAO) threads
($http://cxc.harvard.edu/ciao/guides/index.html$), 
and using the CIAO software package v3.4 (see Massaro et al. 2008b for details).  The Chandra Calibration
Database (CALDB) version 3.4.2 was used to process all files.  
Level 2 event files were generated using the $acis\_process\_events$ task,
after removing the hot pixels with $acis\_run\_hotpix$.  Events were
filtered for grades 0,2,3,4,6 and we removed pixel randomization.
Astrometric registration was done changing the appropiate keywords in
the fits header so as to align the nuclear X-ray position with that of
the radio.  We also registered the HST images in the same way.

We created 3 different fluxmaps (soft, medium, hard, in the ranges 0.5
-- 1, 1 -- 2, 2 -- 7 keV, respectively) by dividing the data with the
exposure maps. When constructing the fluxmaps, we normalized each
count by multiplying by h$\nu$ where $\nu$ corresponds to the energy
used for the corresponding exposure map.  Thus we can measure the flux
in any aperture in cgs units with only a small correction for the
ratio of the mean energy of the counts within the aperture to the
nominal energy for the band.

The total counts detected in the X-ray band between 0.5--7.0 keV is 148, for an extraction region of 6 arcsec centered at the position of the radio nucleus (RA: 14 49 21.661, DEC: +63 16 14.12).
Only  8\% (13 counts) of this counts are in the hard X-ray band (2.0--7.0 keV) while the majority of the X-ray emission lies in the soft X-ray energy range between 0.5--2.0 keV.
Its low flux together with the absence of broad emission lines in the optical spectrum 
suggests that it is absorbed as observed in other type II CSS sources (Siemiginowska et al. 2008). 
The radio map at 1.7 GHz with the soft X-ray contours overlaid is shown in Fig. 1a. 

The soft X-ray emission extends beyond the radio one and the X-ray flux peaks are not coincident with those in the radio.
Comparing the \chn~image with the optical one (see Fig. 3), we found that the X-ray emission is associated with the optical emission line region dominated by the [O III]5007. The coincidence between the soft X-ray and the [O III]5007 is notable, both in extension and in the overall morhpology.
It has also shown in Fig. 1b that the majority of the soft X-ray emission lies in the northern region, this fact is probably due to the presence of the dust lane in the southern region. The number of counts in the southern region is about 2/3 that in the northern one.

The striking resemblance of [O III]5007 structures with the soft X-ray emission favors a common origin for both components.
On the basis of this association, assuming that the soft X-ray radiation is dominated by emission lines,
we selected an extraction region for the soft X-ray emission which covers the [O III]5007 pattern.

We measured the soft X-ray flux in the 0.5-2.0 keV energy range (observer frame) and its ratio with the [O III]5007. 
We found a soft X-ray flux F$_{0.5-2~keV}$ $\sim$3.49$\times$10$^{-14}$ erg~s$^{-1}$~cm$^{-2}$, 
that corresponds to an observed  soft X-ray luminosity L$_{0.5-2~keV}$ $\sim$ 1.38$\times$10$^{41}$ erg~s$^{-1}$ corrected for the Galactic HI absorption (N$_H\simeq 3.30\times 10^{20}~cm^{2}$, Kalberla et al. 2005, assuming a photon index of 2).
The observed [O III]5007 flux is F$_{[O III]}$ $\sim$3.68$\times$10$^{-14}$ erg~s$^{-1}$~cm$^{-2}$ corrected for the Galactic extinction (A$_{\lambda 5007}$ $\simeq$0.097, Cardelli et al. 1989).
Their ratio F$_[O III]$/ F$_{0.5-2~keV}$ is of the order unity.

The \xmm~spectrum available in the public archive does not allow to discriminate between collisional and photoionized gas.
Evans et al. (2008) investigated the \xmm~spectrum of 3C 305 and they argued that 
it cannot be described by a simple unabsorbed power-law model. They found an acceptable fit 
combining a single unabsorbed power-law with a thermal emission due to collisional ionized gas (APEC model in {\sc xspec} Arnaud 1996).
However, we argue that the statistics of the \xmm~observation cannot exclude a photoionized model to interpret the X-ray emission of 3C 305.

\section{Discussion and Conclusions}
The interpretation of the X-ray extended emission in CSS is still unclear and several scenarios have been proposed to understand its nature.

3C 305 represents the first clear case of a CSS source where the extended X-ray emission is not connected with radio structure. 
Investigating the radio structure of 3C 305 we found that the radio emission at 8.4 GHz is in agreement with that at 1.7 GHz, 
showing that the jets, lobes, and also the two bright spots have a moderately steep spectrum. 
This fact together with the small size extension suggests that 3C 305 belongs to the class of Compact Steep Spectrum radio galaxies. 

On the basis of our morphological study, we showed that the extended soft X-ray emission lies beyond the radio one. The X-ray flux peaks are not coincident with those in the radio mostly correspondent to the hotspots.
This fact led us to conclude that the X-rays are not produced by the same mechanism responsible of the radio jet emission.

The pattern of the X-ray radiation compared with the radio jet emission seems to support the scenario of a collisional ionized gas driven by shocks, in which the bow shock is produced beyond the radio emission. 
On the other hand, by comparing the \chn~image with the optical one, we found that the X-ray emission is co-spatial with the optical emission line region dominated by the [O III]5007.
This favors the idea that the origin of the X-ray emission has thermal nature, and that it could be interpreted as due 
to the interaction between the radio jet and the ISM. 
It is worth noting that the majority of the X-ray flux is below 2 keV, where the majority of emission lines of collisional ionized gas are present.

However, the association between the X-ray and the [O III]5007 region and the X-ray emission itself can be also described 
as due to photoionized gas, following the same interpretation as for the Seyfert galaxies. Thus soft X-ray energy range could be dominated emission lines due to photoionized gas.

Considerable debate has been developed on the interpretation of these extended structures in Seyfert galaxies, 
where it has been associated with the optical emission line region, dominated by the [O III]5007.
Two different scenarios have been proposed for this emission: as produced by the radio jet interactions with the circumnuclear environment, 
(collisional ionization or a radiative ÕautoionizingÕ shock as argued by Dopita \&
Sutherland 1995, 1996) or by emission from photoionized gas, in the form of blend of emission lines (as claimed by Bianchi et al. 2006).
Recently, the latter scenario appears to be favored on the basis of the comparison between 
the \chn~ X-ray observations and the HST images (Bianchi et al. 2006). 
Our recent \chn~observations of 3C 305, presented in this work, shows that this debate can be extended to the CSS radio galaxies.

We evaluated the ratio between the [O III]5007 and the soft X-ray flux and we found that it is of the order unity, 
the same order of magnitude of Seyfert galaxies by Bianchi et al. (2006). Furthermore, we note that
Simpson et al. (1996) have presented a diagnostic diagram involving the ratio [O I]/H$\alpha$ to study the emission line region, 
and 3C 305 lies approximately in the region of Seyfert galaxies in this diagram (Jackson et al. 2003). 
Such behavior is in agreement with our results for the ([O III]5007/F$_{0.5-2~keV}$) ratio.
This analogies with the case of Seyfert galaxies suggest that the nature of the extended X-ray emission could have the same origin.

Even if there are some similarities with the Seyfert galaxies scenario, the past preferred interpretation of the extended emission line region in 3C 305 is in favor of the interaction
between the radio jet and the the ISM (Heckman et al. 1982, Jackson et al. 1995, 2003). 3C 305 represents a unique case to investigate this process because
with respect to the case of Seyfert galaxies it is a powerful radio source.

The discovery of the association between the soft X-ray emission and the [O III]5007 emission line region in 3C 305 constitutes a first step to understand the nature of the extended emission line region in CSS sources. Our results strongly suggest that in this case the extended X-ray emission has a thermal origin. 
However the short \chn~observation does not allow us to solve the dichotomy between a jet-driven and a photoionization scenarios. An investigation based on a deeper \xmm~ spectral analysis and a deeper \chn~ observation is crucial to achieve the final interpretation.
 
\smallskip
We thank D. E. Harris for his suggestions and for several comments and
G. Privon for the HST images.
F. Massaro is grateful to S. Bianchi for his suggestions, to M. Murgia for discussions on radio reduction 
and to A. Siemiginowska for comments on X-ray data analysis.  
F. Massaro, A. Bonafede and E. Liuzzo thanks E. Bressert for his help to 
present the image of 3C 305 at the "NRAO image contest 2008" (2nd prize).
Finally, we thank our referee for several comments helpful toward improving our presentation.
This research has made use of NASA's Astrophysics Data System and SAOImage
DS9, developed by the Smithsonian Astrophysical Observatory and the NASA/IPAC Extragalactic Database (NED) which is
operated by the Jet Propulsion Laboratory, California Institute of
Technology, under contract with the National Aeronautics and Space
Administration.  
The National Radio Astronomy Observatory is operated by Associated 
Universities, Inc., under contract with the National Science Foundation.
The work at SAO is supported by NASA-GRANT GO8-9114A.



\clearpage

\end{document}